\documentclass{Interspeech}

\interspeechcameraready

\usepackage{xcolor}

\usepackage{amsmath,amsfonts,bm}

\def\eqref#1{equation~\ref{#1}}

\def\1{\bm{1}}

\newcommand{\bld}[1]{\boldsymbol{#1}}

\def\rvf{{\mathbf{f}}}

\def\rvk{{\mathbf{k}}}

\def\rvx{{\mathbf{x}}}
\def\rvy{{\mathbf{y}}}
\def\rvz{{\mathbf{z}}}

\def\rmK{{\mathbf{K}}}

\def\rmX{{\mathbf{X}}}

\DeclareMathAlphabet{\mathsfit}{\encodingdefault}{\sfdefault}{m}{sl}
\SetMathAlphabet{\mathsfit}{bold}{\encodingdefault}{\sfdefault}{bx}{n}

\newcommand{\rvmu}{\bld{\mathbf{\mu}}}

\newcommand\ignore[1]{{}}
\definecolor{darkpink}{RGB}{180, 0, 90} 

\title{Few-Shot Speech Deepfake Detection Adaptation with Gaussian Processes}

\author[affiliation={}]{Neta}{Glazer}
\author[affiliation={}]{David}{Chernin}
\author[affiliation={}]{Idan}{Achituve}
\author[affiliation={}]{Sharon}{Gannot}
\author[affiliation={}]{Ethan}{Fetaya}
\usepackage{algorithm}
\usepackage{algpseudocode}
\usepackage{placeins}
\usepackage{hyperref}

\affiliation{}{Bar Ilan University}{Israel}
\email{netaglazer@gmail.com, ethan.fetaya@biu.ac.il}

\usepackage{comment}

\begin{document}

\maketitle

\begin{abstract}
    
Recent advancements in Text-to-Speech (TTS) models, particularly in voice cloning, have intensified the demand for adaptable and efficient deepfake detection methods. As TTS systems continue to evolve, detection models must be able to efficiently adapt to previously unseen generation models with minimal data. This paper introduces ADD-GP, a few-shot adaptive framework based on a Gaussian Process (GP) classifier for Audio Deepfake Detection (ADD). We show how the combination of a powerful deep embedding model with the Gaussian processes flexibility can achieve strong performance and adaptability.
Additionally, we show this approach can also be used for personalized detection, with greater robustness to new TTS models and one-shot adaptability. To support our evaluation, a benchmark dataset is constructed for this task using new state-of-the-art voice cloning models. 
\end{abstract}

\section{Introduction}

The rapid advancements in text-to-speech (TTS) generation have significantly improved the ability to synthesize highly realistic speech from textual input \cite{wang2023neural, le2024voicebox}. Recent models also excel in voice cloning, enabling the generation of speech that closely mimics a specific speaker's voice with just a single reference utterance \cite{casanova2022yourtts, chen2024f5}. This breakthrough has significantly enhanced applications in accessibility, entertainment, and human-computer interaction.

However, these advancements also introduce significant security and ethical concerns, particularly in the realm of voice cloning. With only a short audio sample, modern voice cloning models can generate speech that is nearly indistinguishable from the original speaker’s voice. This capability has been exploited in various malicious activities, including fraud, misinformation, and identity theft. 
For instance, fraudsters have successfully mimicked corporate executives to authorize fraudulent financial transactions, leading to substantial financial losses. Similarly, scammers have used synthetic voices to pose as family members in distress, coercing victims into transferring money under false pretenses \cite{shaaban2023audio}.

This problem is addressed by Audio Deepfake Detection (ADD) algorithms. We observed that current ADD detectors easily detect deepfakes from known TTS models. However, they may generalize poorly to deepfakes from unseen generation models. This is a significant security concern as new models are introduced at an accelerating rate. Moreover, advancement in parameter efficient fine-tuning, e.g., LoRA \cite{hulora}, allows attackers to modify an existing TTS model on a relatively small amount of data. To address this challenge,  a detection system must be capable of efficiently adapting to novel deepfake generation methods with minimal data, allowing it to continuously evolve against new threats. We also note that many leading commercial TTS models, such as 11Labs, are only accessible via paid APIs. This limits the feasibility of large-scale data collection for retraining, further emphasizing the need for efficient few-shot adaptation.

In this paper, we introduce a new system for adaptive deepfake detection based on Gaussian Processes (GP), which we name ADD-GP. We use a Deep Kernel Learning (DKL) framework,  based on XLS-R features \cite{babu2021xls}, to improve the generalization capability of detection models. Gaussian Processes (GPs) are well-suited for few-shot detection due to their non-parametric nature and ability to provide uncertainty estimates. By integrating GPs with deep neural networks through Deep Kernel Learning (DKL), we enable the model to learn highly expressive feature representations while maintaining the flexibility and robustness of GPs. Finally, we explore the use of our framework for personalized deepfake detection. This shows improved robustness to new TTS models and better adaptability even in cases where only a single generated example is available.

Our key contributions include:
\begin{itemize}
\item A novel method for audio deepfake detection, enabling effective few-shot adaptation for new TTSs.
\item A state-of-the-art personalized few-shot deepfake detection with greater robustness and adaptability.
\item The introduction of \textbf{LibriFake}, a dataset specifically designed for evaluating few-shot and one-shot deepfake audio detection approaches on novel TTS models. To support future research and ensure the reproducibility of our results, we have made both our source code and datasets publicly available at: \textcolor{magenta}{\url{https://github.com/NetaGlazer/ADD-GP}}
\end{itemize}

\section{Background}
We provide a brief introduction to Gaussian Process (GP) classifiers, which form the foundation of our proposed method. Scalars are denoted with lowercase letters (e.g., $x$), vectors with boldface lowercase letters (e.g., $\mathbf{x}$), and matrices with boldface uppercase letters (e.g., $\mathbf{X}$). In general, $\mathbf{y} = [y_1, ..., y_N]^T$ represents the vector of labels, and $\mathbf{X} \in \mathbb{R}^{N \times d}$ represents the design matrix comprising $N$ data points of dimension $d$, where the $i$-th row is $\mathbf{x}_i$.

\subsection{Gaussian Processes}
Gaussian Processes (GPs) define a distribution over functions. 
A GP model is fully specified by a mean function $m(\rvx)$ and a covariance function (kernel) $k(\rvx, \rvx')$. We say that $\rvf \sim \mathcal{GP}(m(\rvx), k(\rvx, \rvx'))$ if for a given dataset $\mathbf{X}$, the function values $\mathbf{f} = [f(\rvx_1), ..., f(\rvx_N)]^T$ are normally distributed, i.e., $\mathbf{f} \sim \mathcal{N}(\boldsymbol{\mu}, \mathbf{K})$, where $\rvmu_i = m(\rvx_i)$ and $\rmK_{ij} = k(\rvx_i, \rvx_j)$~\cite{gp_book}. Commonly, the mean function is the constant zero function. The predictive power of GPs mainly comes from the kernel which encodes similarity between the data points. For regression tasks if the likelihood $p(\rvy|\rvf,\mathbf{X})$ takes a Gaussian form, 
the posterior distribution $p(\rvf|\rvy,\mathbf{X})$ is a Gaussian with a closed-form formula for the mean and variance. At inference, given a new data point $\rvx_*$, the predictive distribution $p(\rvf_*|\rvx_*,\rvy,\mathbf{X})=\int p(\rvf_*|\rvx_*,\rvf)p(\rvf|\rvy,\mathbf{X})d\rvf$ is also a Gaussian with parameters given by:
\begin{align}
    &p(\rvf_*|\rvx_*,\rvy,\mathbf{X})=\mathcal{N}(\mu_*,\sigma_*^2) \\
    &\mu_*=\rvk_*^T(\mathbf{K}+\sigma^2\mathbf{I})^{-1}\rvy\\
    &\sigma_*=k_{**}-\rvk_*^T(\mathbf{K}+\sigma^2\mathbf{I})^{-1}\rvk_*
\end{align}
where $k_{**}=k(\rvx_*,\rvx_*)$, and $\rvk_*[i]=k(\rvx_i,\rvx_*)$.  This means that given a kernel and a dataset, the GP prediction is given by a closed-form solution without any optimization. \\

\subsection{Gaussian Process Classification (GPC)}
\label{sec:GPC}
Unlike GP regression with a Gaussian likelihood, in Gaussian Process Classification (GPC), the likelihood $p(\mathbf{y} | \mathbf{f}, \rmX)$ is non-Gaussian, making exact posterior inference intractable. To address this, approximate inference techniques, such as variational inference (VI) \cite{hensman2015scalable, wilson2016stochastic, achituve2023guided}, or Markov chain Monte Carlo (MCMC) \cite{polya_gamma, galy2020multi, achituve2021gp, achituve2021personalized}, are commonly used. Here, we follow the approach presented in \cite{milios2018dirichlet}, although our approach can be combined with any standard GP classifier. It is important to note that this approach still uses a closed-form solution for the posterior and at inference.  

\subsection{Deep Kernel Learning (DKL)}
\label{sec:DKL}
The performance of GP models is heavily dependent on the choice of kernel function $k(\rvx_i, \rvx_j)$ with standard kernels often failing to capture meaningful semantic similarities in complex data modalities such as images and speech. Deep Kernel Learning (DKL) addresses this limitation by combining classical kernels with a neural network transformation \cite{calandra2016manifold, gordon16_DKL}. For example, it is common to use the RBF kernel  on features computed by a deep network $g_\theta$:
\begin{equation}\label{eq:rbf}
k_\theta(\rvx_i, \rvx_j) = \sigma \cdot \exp\left(- \frac{||g_\theta(\rvx_i) - g_\theta(\rvx_j)||^2}{2\ell^2}\right).
\end{equation}
Here, the kernel parameters are (1) the neural network parameters $\theta$, (2) the length scale $\ell$ and (3) the output scale $\sigma$. Typically, all the parameters are learned using a Bayesian loss, such as the log marginal likelihood or the evidence lower bound (ELBO).  

\subsection{Audio Deepfake Detection}

Audio Deepfake Detection (ADD) methods generally follow two main architectures: pipeline-based and end-to-end (E2E) based approach. In pipeline-based models \cite{yi2023audio}, the system is split into two distinct components: a front-end feature extractor and a back-end classifier \cite{Jung2021AASIST}. The front-end extracts features such as Mel-Frequency Cepstral Coefficients (MFCC), Linear Frequency Cepstral Coefficients (LFCC), or embeddings from a deep neural network, e.g., wav2vec2, which are then passed to an independent back-end classifier for classification.

Some E2E models, like SSL-AASIST \cite{jung2022aasist} still use pretrained feature extractors, but instead of keeping them fixed, they fine-tune both the front-end feature extractor and the back-end classifier together. Other E2E models, such as RawGAT-ST \cite{tak21_asvspoof} and RawNet3 \cite{jung2022pushing}, learn directly from raw waveforms, extracting features and classifying them within the same model without a separate feature extraction step.

Wav2Vec2 \cite{schneider2019wav2vec}, HuBERT \cite{hsu2021hubert}, and XLS-R \cite{babu2021xls} have become standard backbone models in modern ADD systems due to their strong ability to extract meaningful audio representations \cite{li2024cross}. These speech foundation models are widely used in both pipeline-based and end-to-end architectures, either as fixed feature extractors or fine-tuned alongside classifiers. Compared to handcrafted features and traditional deepfake detection models, self-supervised speech models offer stronger generalizability, especially when tested on unseen datasets \cite{li2024sonar}. 

\begin{table*}[t]
\centering
\caption{Evaluation of 11Labs using Wav2Vec2-2B. Equal Error Rate (EER) in percentage averaged over three random seeds.}
\label{tab:11labs_evaluation}
\renewcommand{\arraystretch}{1.6} 
\fontsize{11pt}{13pt}\selectfont  
\resizebox{\textwidth}{!}{%
\begin{tabular}{lccccccccc}
    \toprule
    & ID TTS& OOD 11Labs& 5-Shots & 10-Shots & 20-Shots & 50-Shots & 100-Shots & 100-Shots ID & Params \\ 
    \midrule
    SSL-AASIST-FT& $0.2\%$& $42.48\%$ & $16.17$ $\pm$ $0.62$ & $15.55$ $\pm$ $0.01$ & $12.56$ $\pm$ $0.27$ & $11.73$ $\pm$ $0.75$ & $10.97$ $\pm$ $0.40$ & $2.3\%$& 2b\\ 
    SSL-AASIST-FT MixPro& $0.2\%$& $42.48\%$ & $17.44$ $\pm$ $0.09$ & $14.10$ $\pm$ $0.61$ & $12.22$ $\pm$ $0.27$ &	$8.26$ $\pm$ $0.09$ & $5.76$ $\pm$ $0.27$ & 5.42\%& 2b\\  \hline
    
    SSL-AASIST-Lora-FT& $0.2$\% & $42.48$\% & $22.98$ $\pm$ $1.33$ & $20.27$ $\pm$ $0.09$ & $20.62$ $\pm$ $0.39$ & $20.35$ $\pm$ $0.01$ & $20.20$ $\pm$ $0.01$ & $1.02$\%& 5m\\ 
    SSL-AASIST-Lora-FT MixPro & $0.2$\%& $42.48$\% & $24.59$ $\pm$ $1.5$& $21.25$ $\pm$ $0.03$& $21.04$ $\pm$ $0.35$& $19.79$ $\pm$ $0.02$& $19.59$ $\pm$ $0.02$& $1.33$\%& 5m\\ \hline
    OWM & $0.2$\%& $42.48$\% & $26.02$ $\pm$ $1.67$ & $23.97$ $\pm$ $0.04$ & $20.37$ $\pm$ $0.56$ & $11.54$ $\pm$ $0.09$ & $7.85$ $\pm$ $0.16$ & $7.29$\%& 2b\\ 
    OWM MixPro & $0.2$\%& $42.48$\% & $19.00$ $\pm$ $1.57$& $13.02$ $\pm$ $0.08$& $10.12$ $\pm$ $0.34$& $7.85$ $\pm$ $0.07$& $3.93$ $\pm$ $0.15$& $1.42$\%& 2b\\ 
    \midrule
    RWM & $0.2$\%& $42.48$\% & $20.04$ $\pm$ $1.15$ & $16.94$ $\pm$ $0.29$ & $15.98$ $\pm$ $0.39$ & $5.09$ $\pm$ $0.74$ & $4.41$ $\pm$ $0.09$ & $0.51$\%& 2b\\ 
    RWM MixPro & $0.2$\%& $42.48$\% & $14.75$ $\pm$ $0.7$& $12.30$ $\pm$ $0.5$& $6.97$ $\pm$ $0.18$& $4.92$ $\pm$ $0.68$& $1.24$ $\pm$ $0.05$& $0.71$\%& 2b\\ 
   \midrule
        CD-ADD & $0.4$\%& \textbf{$35.12$\%} & $32.37$ $\pm$ $1.63$ & $28.28$ $\pm$ $0.67$ & $15.29$ $\pm$ $0.39$ & $9.98$ $\pm$ $1.31$ & $7.92$ $\pm$ $0.12$ & $23.68$\%& 2b\\ 
    CD-ADD MixPro & $0.4$\%& \textbf{$35.12$\%} & $28.93$ $\pm$ $1.51$& $26.45$ $\pm$ $0.57$& $19.21$ $\pm$ $0.21$& $8.88$ $\pm$ $0.96$& $7.23$ $\pm$ $0.09$& $23.68$\%& 2b\\\bottomrule 
    \textbf{ADD-GP (ours)} & \textbf{$0.1$\%}& $40.31$\% & \textbf{$13.44$} $\pm$ \textbf{$0.01$} & \textbf{$12.80$} $\pm$ \textbf{$0.03$} & \textbf{$8.72$} $\pm$ \textbf{$0.03$} & \textbf{$4.70$} $\pm$ \textbf{$0.04$} & \textbf{$2.78$} $\pm$ \textbf{$0.03$} & \textbf{$0.1$\%} & - \\ 
    \textbf{ADD-GP MixPro (ours)} & \textbf{$0.1$\%}& $40.31$\% & \textbf{$7.89$ $\pm$ $0.01$} & \textbf{$4.86$ $\pm$ $0.01$} &  \textbf{$3.31$ $\pm$ $0.08$} & \textbf{$1.14$ $\pm$   $0.01$} & \textbf{$0.54$ $\pm$ $0.03$}&  \textbf{$0.1$\% }& - \\ 
\bottomrule
\end{tabular}}
\end{table*}

\section{Methods}


We introduce our framework that uses a Dirichlet-based GP classifier \cite{milios2018dirichlet} as a back-end, on features extracted from XLS-R \cite{babu2021xls}, a Wav2Vec2 based self-supervised model, as the front-end.


\subsection{Datasets Generation}
To evaluate our method, we introduce LibriFake, a large-scale deepfake benchmark derived from LibriSpeech 
\cite{panayotov2015librispeech}.  For each sample in LibriSpeech, we generate a corresponding synthetic version using voice cloning with the original text transcription, effectively replicating the dataset for each TTS model. The dataset is then split into training and testing subsets, with $\sim15\%$ of speakers in the test set, ensuring that no samples from the same speaker appear in both sets. This separation is designed to test the model’s ability to generalize to new speakers. We used: (i) yourTTS \cite{casanova2022yourtts}, (ii) Whisper-Speech \cite{SpearTTS}, (iii) Vall-e-x \cite{zhang2023speak},  (iv) F5-TTS \cite{chen2024f5}, and (v) 11Labs.\footnote{https://elevenlabs.io/} When evaluating generalization to unseen TTS, we found that detection performance was significantly worse on 11Labs. Therefore, we selected it as the unseen TTS for our few-shot adaptation experiments. For a personalized out-of-distribution evaluation, we also selected 30 speakers from the VoxCeleb dataset \cite{nagrani2017voxceleb} and generated fake samples from them using the same process we employed to create LibriFake. Because VoxCeleb, unlike LibriSpeech, does not include transcripts, we first transcribed the speakers’ audio with Whisper-Large-v2. These transcriptions were then used for the TTS text conditioning.

\subsection{GP Deepfake Detection}
In the first phase, we train a Gaussian Process (GP) classifier to distinguish between real and fake speech samples from our dataset. We used the training set of all available TTS models, except for the 11Labs model, which was left unseen during training, and 1000 random training examples excluded from the kernel learning phase. We used the RBF kernel on features computed by XLS-R, which was frozen except for the last block. This means that during training, we only update the last XLS-R block, the lengths scale, and the output scale. At each iteration, we draw 80 samples, embed them via XLS-R, build a GP classifier using the RBF-kernel on these features (eq. \ref{eq:rbf}), compute the log-marginal likelihood, backpropagate, and update the kernel parameters. During the evaluation phase, we fixed our tuned parameters and built a GP classifier using our learned kernel on the 1000 training samples we kept aside. We show pseudocode in Alg. \ref{alg:gp_deepfake_detection}.

\subsection{Few-Shot Adaptation}
As the kernel only measures similarity, it can generalize well. As such, we can keep the kernel fixed and only update the GP with the fixed kernel. As GPs are a non-parametric method, adapting them without any training is straightforward. We simply add the examples from the new TTS to the held-out 1000 training examples and build a new GP classifier with these additional examples, i.e. recomputing lines 10-11 in Alg. \ref{alg:gp_deepfake_detection} with an augmented $\rvz_{\text{eval}}$.


\subsection{MixPro for Few-Shot Adaptation}
\label{sec:mixup} 
To further enhance generalization in few-shot scenarios, we incorporate MixPro \cite{xue2023few}, a data augmentation technique based on MixUp \cite{zhang2017mixup}, specifically designed for few-shot domain adaptation. MixPro generates additional training examples by blending source and target domain samples in the embedding space.  

More precisely, given a neural network embedding function \( f \) and two samples \( \rvx_s \) and \( \rvx_t \) from the same class—one from the source domain and one from the target domain—we create mixed examples using the following interpolation:  
\begin{equation}
f_{\text{mix}} = (1 - \lambda) f(\rvx_s) + \lambda f(\rvx_t),
\end{equation}
where \( \lambda \) is a random interpolation parameter. In our case, only the fake samples are OOD, so we only mix deepfake samples from the new TTS with deepfake samples from the training TTSs.
 In our GP few-shot adaptation experiments, we created an augmented dataset that was 20 times larger than the original and combined it with the original samples. For the baseline methods, we iteratively sampled augmented examples at each fine-tuning step.


\begin{table*}[!h]
\centering
\caption{Personalized evaluation on LibriSpeech (LS), and VoxCeleb (VC) subsets.}
\renewcommand{\arraystretch}{1.2} 
\begin{tabular}{l lcllcl}
    \toprule
    &   OOD (LS)&1-shot (LS) &5-shot (LS)& OOD  (VC)& 1-shot (VC)  &5-shot (VC)\\
    \midrule
    SSL-ASSIST-FT&  $5.92\% $ & $5.18\%$&$4.13\%$&  $11.07\%$& $10.4\%$&$6.41\%$\\
    SSL-ASSIST-LORA-FT&    $19.89\%$& $18.33\%$&$15.81\%$& $48.01\%$& $46.42\%$&$46.42\%$\\
    CD-ADD&    $11.24\%$& $6.65\%$& $3.69\% $& $10.03\%$& $9.76\%$&$9.17\%$\\
 OWM& $13.25\%$ & $9.04\%$ & $7.23\%$ & $10.86\%$& $9.11\%$&$6.34\%$\\
    RWM &    $7.75\%$ & $7.1\%$ & $6.06\%$& $12.8\%$& $7.39\%$&$6.69\%$\\ \midrule
    ADD-GP (ours)&    $\textbf{2.59\%}$ & $\textbf{0.61\%}$ & $\textbf{0.39\%}$& $\textbf{6.95\%}$& $\textbf{2.03\%}$& $\textbf{0.52\%}$ \\
    \bottomrule
    
    \end{tabular}
\label{tab:personal}
\end{table*}

\section{Experiments}

\subsection{Baselines}

We evaluated our approach against SSL-AASIST \cite{tak2022automatic}, a widely used baseline known for its strong performance on publicly available datasets. SSL-AASIST employs the AASIST \cite{jung2022aasist} architecture as the back-end while using XLS-R as the front-end, making it a relevant benchmark for comparison with our XLS-R-based framework.

To adapt SSL-AASIST to few-shot learning, we fine-tuned the model both with and without LoRA, as well as using Orthogonal Weight Matching (OWM) \cite{zeng2019continual}, a continual learning method designed to mitigate catastrophic forgetting. Additionally, we incorporated Remember Weight Matching (RWM) \cite{zhang2024remember}, a recently introduced continual learning approach that has demonstrated strong few-shot adaptation capabilities. Notably, RWM was originally implemented with an XLS-R backbone, aligning closely with our setup.
Furthermore, we implemented the CD-ADD method \cite{li2024cross}, which has also been shown to support few-shot adaptation. To ensure a fair comparison, all baselines were trained and evaluated using the same XLS-R parameters.

\subsection{Few-Shot Adaptation for New TTS Models}\label{sec:man_exp}
We first trained all baselines on our LibriFake training datasets, including all TTS models except for 11Labs, which will be used as the out-of-distribution TTS. We then adapted each baseline with 5, 10, 20, 50, and 100 samples from 11Labs. We also train each method on 11Labs samples using the MixPro method, explained in Sec.~\ref{sec:mixup}.  
We run each adaptation method with three different random seeds and present the average results. 

Table~\ref{tab:11labs_evaluation} presents the equal error rate (EER) performance across different settings. First, in the In-Distribution (ID) TTS column, we report the EER on test data from TTS systems seen during training. As expected, all methods perform exceptionally well in this scenario. Next, in the Out-Of-Distribution (OOD) 11Labs setting, we evaluate how well the original models detect deepfake audio from the unseen 11Labs TTS on test speakers. The results indicate that all methods struggle with generalization, leading to poor performance on this task. We then examine $k$-shot adaptation, where our ADD-GP method demonstrates a significant advantage over all baselines. Additionally, we observe improvement across most methods when incorporating MixPro augmentation. Finally, we test how the model adapted on 100-shot 11Labs samples performs on the original TTS seen during training. All methods, except for our ADD-GP, show a decrease in performance on the original TTSs, even when using methods designed to mitigate ``catastrophic forgetting''. We note that this aligns with previous works that showed good continual learning results with GP models \cite{achituve2021gp}.

\subsection{Personalized Deepfake Detection}
We also explored personalized deepfake detection to highlight the additional benefits of the GP approach’s flexibility. In this experiment, each method was tailored to detect deepfakes for a specific speaker. For our GP classifier, we used the learned kernel with 20 samples—10 real and 10 fake—from a single individual to create a personalized detector. For the baseline methods, we fine-tuned them on the same 20 samples. To adapt to a new, unseen TTS, we used samples generated from that TTS for our personalized speaker, following the same approach as in Sec. \ref{sec:man_exp} with MixPro.
We show our results in Table. \ref{tab:personal} for test speakers from LibreFake dataset, as well as new speakers taken from VoxCeleb \cite{nagrani2017voxceleb}. Each experiment was repeated 30 times on 30 different speakers. 

As one can see, in general, all personalized methods generalize better to the unseen 11Labs TTS samples when compared to their general detector counterpart. For our ADD-GP method, we achieve $0.61\%$ EER with a 1-shot adaptation on LibriFake, comparable to a 100-shot adaptation in the non-personalized experiment. It is important to note that while this personalized detector exhibits excellent performance, it is limited by the fact that it requires deepfake samples specifically for the target speaker rather than relying on generic deepfake samples.

\begin{figure}[t]
    \centering
    \captionsetup{justification=centering} 

    \begin{minipage}{0.52\linewidth}
        \centering
        \includegraphics[width=0.98\linewidth]{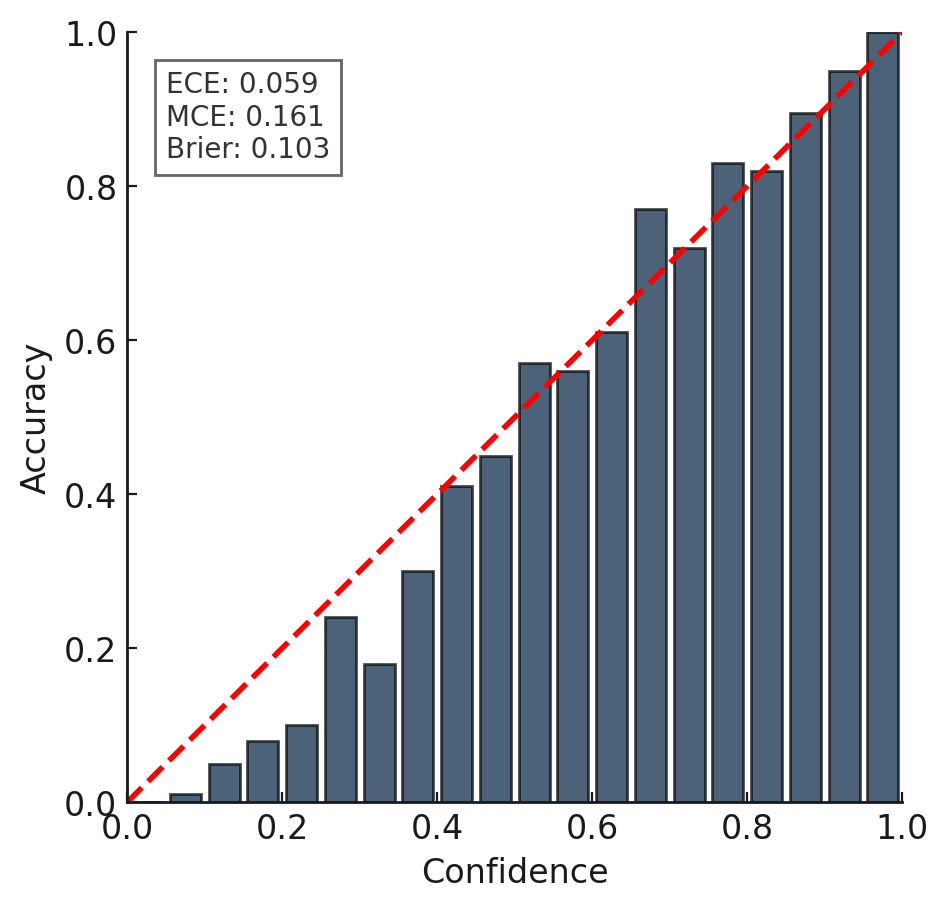}
    \end{minipage}
    \hspace{-0.65cm} 
    \begin{minipage}{0.52\linewidth}
        \centering
        \includegraphics[width=0.98\linewidth]{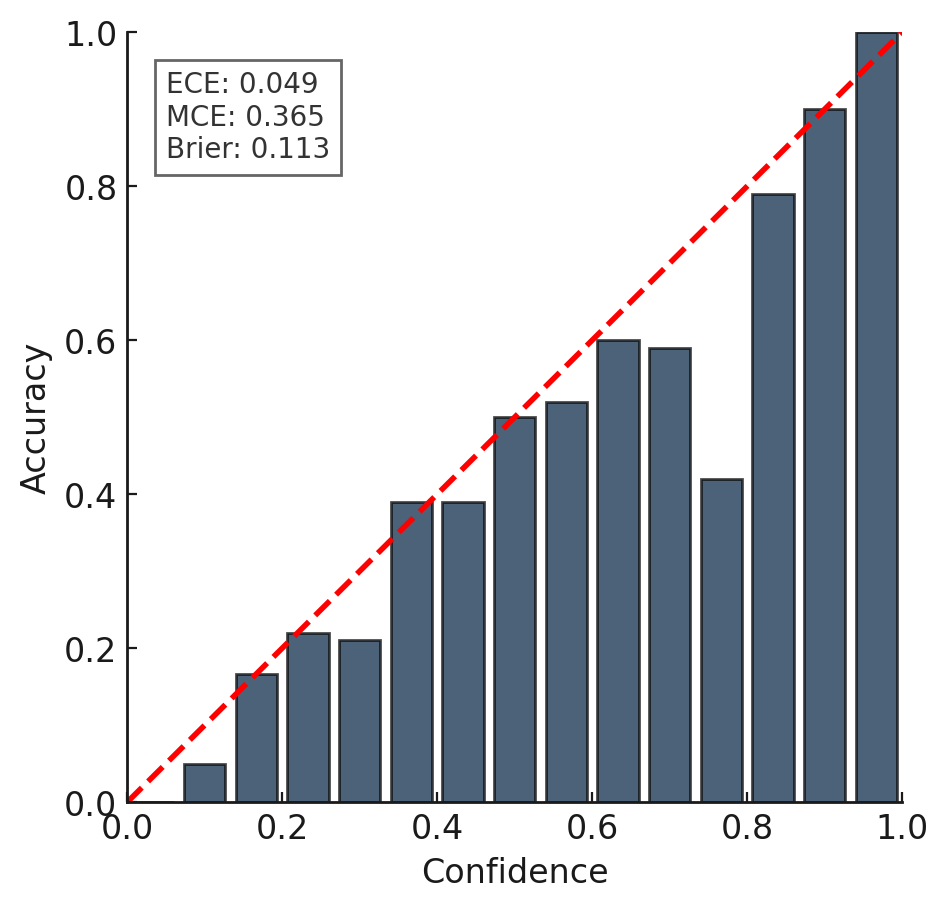}
    \end{minipage}
    \caption{Calibration curves for different models. On the left is the calibration on the 10-shot experiment. On the right, calibration for personalized 5-shot VoxCeleb experiment.}
    \vspace{-0.3cm}
    \label{fig:calibration_curves}
\end{figure}

\subsection{Calibration}
An additional benefit of the Bayesian GP approach is that it offers an accurate measure of uncertainty, thereby providing users with a reliable measure of confidence. This is particularly advantageous for ADD, as it informs users of the model’s confidence rather than receiving a simple binary decision. One way to measure uncertainty estimation is via calibration, where we see how the returned probabilities relate to accuracy. The predictions are binned according to $\max_\rvy p(\rvy|\rvx)$, and each bin's average probability is compared to the average accuracy of these samples. As shown in Fig.~\ref{fig:calibration_curves}, the ADD-GP returns well-calibrated results. It is important to note that the SSL-ASSIST and CD-ADD baselines return a score and not a probability and, as such, cannot be evaluated using the same calibration metrics.

\FloatBarrier 
\begin{algorithm}[H]
\small 
\caption{GP Deepfake Detection}
\label{alg:gp_deepfake_detection}
\begin{algorithmic}[1]
\Require Dataset $\mathcal{D} = \{(\rvx_i, y_i)\}_{i=1}^{N}$, divide to $\mathcal{D}_{train}$, $\mathcal{D}_{eval}$, where  $\rvx_i$ are speech samples and $y_i$ are labels. A GP classifier with RBF kernel, Pretrained XLS-R model (only last block trainable).
\State \textbf{Training Phase:}
\For{$t = 1$ to $T$}
    \State Sample batch of 80 examples from $\mathcal{D}$
    \State Compute embeddings $\mathbf{z} = f_{\text{XLS-R}, \theta}(\rvx)$ using XLS-R
    
    \State Fit a GP classifier using RBF kernel:
    \[
        k(\mathbf{z}_i, \mathbf{z}_j) = \sigma^2 \exp\left(-\frac{\Vert\mathbf{z}_i - \mathbf{z}_j\Vert^2}{2\ell^2}\right)
    \]
    \State Compute log-marginal likelihood $\mathcal{L}$
    \State Compute gradients $\nabla\mathcal{L}$ and update trainable XLS-R parameters $\theta$, the length scale $\ell$ and the output scale $\sigma$
\EndFor
\State \textbf{Evaluation Phase:}
\State Load the 1000 reserved training examples $\mathcal{D}_{\text{eval}}$, and compute embeddings $\mathbf{z}_{\text{eval}} = f_{\text{XLS-R}, \theta^*}(\rvx_{\text{eval}})$
\State Predict using $p(\rvf_*|\rvz_*,\rvy_{\text{eval}},\mathbf{z}_{\text{eval}})$
\State Compute classification metrics on the test set

\end{algorithmic}
\end{algorithm}
\FloatBarrier 

\section{Conclusion}
In this work, we address the issue of adapting audio deepfake detection to new unseen TTSs. As demonstrated by our experiments, this can be a significant security concern, as ADD performance on seen TTS is almost perfect while degrading considerably on unseen TTSs. We tackle this issue by introducing a novel ADD approach based on Gaussian processes, which we name ADD-GP. Our experiments show that this model can adapt better to new TTSs with relatively few samples. It also has additional benefits, such as better personalization and returning well-calibrated predictions. This shows how this can be an important tool to address the dangers of deepfake-based fraud in a dynamic and evolving environment. 


\bibliographystyle{IEEEtran}
\bibliography{mybib}

\end{document}